\documentclass[aps,amsfonts,preprint,nofootinbib]{revtex4}
\def\S2{\bar{S}}

\def\d{\delta}

\def\and{a_{n}^\dagger}

\def\sn2d{\Sn2^\dagger}

\def\({\left(}
\def\){\right)}
\def\<{\left\langle}
\def\>{\right\rangle}

\def\d{\nabla}
\def\e{\epsilon_{\mu\nu\rho}}

\newcommand\ee{\end{eqnarray}}      
\newcommand\be{\begin{eqnarray}}
\newcommand\ba{\begin{array}}           
\newcommand\ea{\end{array}}
\newcommand\eeq{\end{equation}}     
\newcommand\beq{\begin{equation}}
\usepackage{amsmath}

\begin{document}
\title{Quantization of Open strings in time dependent Black Holes}
\author{D\'afni F. Z. Marchioro, Daniel L. Nedel}

\affiliation{Universidade Federal da Integra\c{c}\~ao Latino-Americana \\
Avenida Tancredo Neves 6731, Foz do Igua\c{c}u, Brasil}

\email{dafni.marchioro@unila.edu.br; daniel.nedel@unila.edu.br}

\begin{abstract}
    In this letter, the open string is quantized in a time dependent black hole background. The geometry is defined through an adiabatic approximation of the Vaydia metric. The worldsheet two-point function is derived and it is shown to have the same type of singularity as the flat space one. However, the equal times two-point function depends on the particular Cauchy surface where the worldsheet fields are defined.  Finite temperature effects are incorporated through the Liouville-von Neumann approach to non equilibrium thermodynamics.
\end{abstract}
\maketitle

\section{Introduction}
An outstanding challenge of high energy physics involves the consistent incorporation of quantum phenomena in non trivial gravitational backgrounds, particularly in geometries that do not have timelike Killing vectors. The quantization of fields in non trivial geometries has taught us a lot about the quantum field theory itself and the same should happen in first quantized string theory. From the pioneering work of Vega, Medrano and Sanchez in \cite{Vega} to the recent works in the AdS/CFT context \cite{boer}, the quantization of the string in non trivial backgrounds has shown results of great interest for high energy physics, in particular for a better understanding of the string sigma model. Regarding time dependent backgrounds with singularities, progress was made using time dependent plane wave backgrounds \cite{ varios,nos1,nos2}, where it was studied string mode creation, time dependent entropy production and the behaviour of the string at the cosmological singularities. In \cite{PRT} it was studied a model that admits a pre-Big Bang phase scenario and it was argued that the string passes through the null singular point. In \cite{Marchioro:2020qub} it was shown that the left/right entanglement entropy of the Green-Schwarz superstring  is finite in the cosmological singularity of the background  and hence it is a good measure to explore string theory near singularities. In addition, it was shown that the time dependent left/right closed string entropy  has an equilibrium point close to the singularity and, actually at the singularity, it is equal to the thermodynamic entropy of an open string. Concerning black hole backgrounds, we would like to highlight the work of \cite{boer}, where the open string is quantized in a BTZ black hole background. In this model, the bosonic string is hanging from the boundary of the AdS space and dipping into the horizon. So, the end of the string at the boundary undergoes a random motion because of the Hawking radiation of the transverse fluctuation modes, and it is identified with an external particle in the CFT boundary exhibiting Brownian motion. 

Despite all the progress that has been made in singular time dependent plane wave backgrounds and the rich time dependent sigma model that comes from this type of background, these geometries have timelike Killing vectors. In the present work, we are going to investigate a more involved situation: here, the open string is quantized in a time dependent BTZ black hole, which is achieved from an adiabatic approximation of the Vaydia geometry. By considering only small fluctuations of the worldsheet field, only quadratic terms in the Nambu-Goto action are taken into account. Using this approximation, the open string's equation of motion is resolved in the limit $\displaystyle\frac{r_h}{l} <<1$, where $r_h$ is the horizon position at $t=0$ and $l$ is the AdS radius. In this limit the canonical quantization is carried out for worldsheet fields localized in a particular Cauchy surface. The boundary conditions are chosen in such a way that an end of the open string is attached to a brane that coincides with the position of the horizon at $t = 0$. So, at $t=0$, we have a scenario similar to the one of \cite{boer}. 

In the model studied in \cite{boer}, the other end of the string is placed on the boundary of the AdS, requiring a regularization in the definition of the string's mass. Here we use a more standard parametrization of the fundamental string in such a way that the spatial coordinate $\sigma$ belongs to the interval $\sigma \in [\rho_h,\rho_h+1]$, where $\rho_h$ is the dimesionless horizon position.  We have the following picture: at $t = 0$, the open string is hanging between the horizon $\rho_h$ and a brane located at $\rho_h + 1$; as time goes by, the horizon grows adiabatically so that the string is hanging between points inside and outside the horizon. In this process, we are not taking into account the interaction of the string with the thermal gas of closed strings of the black hole background. This can be done because, for the background used here, the dilaton is constant, so the string coupling constant can be kept small throughout the process.
  
An important ingredient to use string’s perturbative techniques is the two-point function. To this end, it is necessary to carry out the mode summation for the string two-point function and present it in terms of analytic functions. It is shown here that the singularity's structure of worldsheet two-point function at zero temperature is quite similar to the one calculated for a flat space. However, owing to the fact that there is no timelike Killing vector at the induced worldsheet metric, the equal times two-point function depends on time, which means that it depends on the Cauchy surface where the worldsheet fields are defined.  As the induced worldsheet geometry is the same as that for the time dependent BTZ black hole, the worldsheet fields are thermally excited. Nevertheless, because of the time dependence of the background, we need a non equilibrium approach to take into account the thermal effects. We use the Liouville-von Neumann (LvN) approach \cite{Kim:2000xb, Kim:2001pg, Dodonov}, which states that one can study the evolution of non equilibrium systems with respect to invariants of the system that satisfy the LvN equation. By choosing as invariant the thermal density matrix defined at $t=0$ (when the system is in equilibrium with the Hawking radiation), we show that the thermal non equilibrium two-point function can be written in terms of Theta and Q-Gamma functions.

\section{The Geometry}

The simplest example of a time dependent black hole which is explored in string theory, mainly in the context of AdS/CFT, is the Vaidya geometry \cite{Vaidya1, Hubeny:2006yu, timedependent, covariantEntropy}. It represents the collapse of an idealized radiating star black hole and it has a time dependent horizon defining a time dependent temperature, as it has been shown by several methods \cite{ECVD}. The $D+1$ dimensional Vaidya-AdS spacetime is given in Poincar\'e coordinates as

\begin{equation}
ds^2 = - \left(r^2 - \frac{m(v)}{r^{D-2}}\right) dv^2 + 2ldvdr + \frac{r^2}{l^2} \sum_{i=1}^{D-1} dx_i^2 \:, 
\label{vaydia}
\end{equation}

\noindent where $v$ is the light-cone time and $l$ is the AdS radius. This geometry accommodates two kinds of surface of particular interest: the apparent horizon $r_{ah} = 2m$ and the event horizon $r_{eh}$ \cite{SQW}. The role played by the apparent horizon in AdS/CFT correspondence can be studied, for example, in \cite{timedependent}. Following \cite{Li,XLi}, the event horizon satisfies the null surface condition

\begin{equation}
1 - \frac{2m(v)}{r_{eh}} - 2 \frac{dr_{eh}}{dv} = 0 \:, 
\end{equation}

\noindent which yields 

\begin{equation}
r_{eh} = \frac{2m(v)}{1-\dot{r}_{eh}} \:,
\end{equation}

\noindent where $\dot{r}_{eh} = \displaystyle\frac{dr_{eh}}{dv}$. The corresponding radiation temperature is given by

\begin{equation}
T = \frac{1-\dot{r}_{eh}}{8\pi m(v)} = \frac{1}{r_{eh}} \:.
\end{equation}

\noindent It should be emphasized that, as the location of the event horizon depends on $v$, the shape of the black hole, as well as the radiation temperature, changes with time. Note that, for $\displaystyle\frac{\dot{r}_{eh}}{m(v)} \approx 0$, the system resembles a BTZ black hole with time dependent temperature. More precisely, this can be confirmed using the coordinate transformation

\begin{equation}
v=\frac{t}{l}+\frac{l}{2\sqrt{m}}\ln\frac{r-\sqrt{m}}{r+\sqrt{m}}.
\label{tempoconedeluz}
\end{equation}

\noindent For $m(v)$ constant, this transformation leads the metric to a BTZ black hole in Poincar\'e coordinates \cite{BTZ}. By making the approximation $\displaystyle\frac{\dot{m}}{m}\approx 0$ in the coordinate transformation  (\ref{tempoconedeluz}), the 3-dimensional Vaidya spacetime becomes

\begin{equation}
ds^2 = -\frac{(r^2 - m(t))}{l^2}dt^2+ \frac{l^2}{r^2 - m(t)}dr^2 +\frac{r^2}{l^2} dx^2,
\label{BTZV}
\end{equation}

\noindent which is the BTZ black hole with a time dependent horizon $r_h(t)=m(t)$. In \cite{Lindesay:2006gv} it is shown that this kind of time dependent black hole results in singular behaviour of curvature invariants at the horizon. However,  as we are going to show, the singularities are proportional to $[\dot{m}(t)]^2$, which is zero in the adiabatic approximation. 

Let's calculate Christoffel's symbols. Non-null components are
\begin{eqnarray}
\Gamma_{tt}^t &=&-\frac{1}{2}\frac{\dot{m}}{r^2-m}=-\Gamma_{rt}^r \nonumber\\
\Gamma_{tr}^t&=&\frac{r}{r^2-m}=-\Gamma_{rr}^r \nonumber \\
\Gamma_{tt}^r &=&\frac{r(r^2-m)}{l^4}=-\Gamma_{xx}^r \nonumber \\
\Gamma_{rr}^t &=&\frac{1}{2}\frac{l^4 \dot{m}}{(r^2-m)^3},\:\: \Gamma_{xr}^x=\frac{1}{r}
\end{eqnarray}

\noindent So, the Riemann tensor components are

\begin{eqnarray}
R_{rr}&=& -\frac{2}{r^2-m}-\left(\frac{ \dot{m}}{m}\right)^2\frac{2l^4}{m^2(1-\frac{r^2}{m})^4}
-\frac{l^4\ddot{m}}{2(r^2-m)^3} \nonumber \\
R_{tt}&=&2\frac{r^2-m}{l^4}-\left(\frac{\dot{m}}{m}\right)^2\frac{1}{(1-\frac{r^2}{m})^2}-\frac{1}{2}\frac{\ddot{m}}{r^2-m}\nonumber \\
R_{tr}&=& -\frac{1}{2}\left(\frac{\dot{m}}{m} \right)\frac{1}{r(1-\frac{r^2}{m})}\nonumber \\
R_{xx}&=& -2\frac{r^2}{l^4} \label{rieman}
\end{eqnarray}

\noindent Note that, if we assume a linear model ($\ddot{m}=0)$, the Riemann tensor can be written as
\begin{equation}
    R_{\mu\nu}(r,t)= -2\frac{g_{\mu\nu}(r,t)}{l^2} +\frac{\dot{m}}{m}A_{\mu\nu}(r,t) + \left(\frac{\dot{m}}{m}\right)^2C_{\mu\nu}(r,t);\:\:\: \mu,\nu= t,r,x  \label{radiab}
\end{equation}

\noindent where the components of the matrices $A_{\mu\nu}$ and $C_{\mu\nu}$ can be read from equation (\ref{rieman}) and $g_{\mu\nu}(r,t)$ are defined in (\ref{BTZV}). It is easy to see now that, in the adiabatic approximation limit ($\frac{\dot{m}}{m}\rightarrow 0)$,  this solution can be used to construct a well defined string model. The one loop worldsheet beta function's equations, for a bosonic string propagating in a 3D background with negative cosmological constant $\Lambda=-\frac{2}{l^2}$, are

\begin{eqnarray}
R_{\mu\nu} + 2\d_\mu\d_\nu \phi
      - \frac{1}{4} H_{\mu\lambda\sigma}
       {H_\nu}^{\lambda\sigma} &=&0  \nonumber \\
 \d^\mu (e^{-2\phi} H_{\mu\nu\rho}) & =& 0 \nonumber \\
 4\d^2\phi -4(\d\phi)^2 + \frac{4}{l^2} + R -\frac{1}{ 12} H^2 &=& 0 \label{beta},
 \end{eqnarray}
 
 \noindent where $\phi$ is the dilaton and $H_{\mu\lambda\alpha}$ is the Kalb-Ramond field strength ($H=dB$). If we assume that $R_{\mu\nu}$ is given by equation  (\ref{radiab}) in  the adiabatic limit, the equations (\ref{beta}) are solved by 
 
 \begin{equation}
     \phi=0,\:\:\:H_{\mu\nu\rho}=\frac{2\e}{l} \label{h}
 \end{equation}

\noindent This is the same solution found in \cite{Horowitz:1993jc} for the usual time independent BTZ black hole. In particular, in \cite{Tseytlin:1994sb} it was shown that, for this kind of solution, the leading-order equations remain exact to all orders in $\alpha'$.

\section{The string Action }

The model consists of an open string stretching between $r = r_{h}$, the horizon position at $t=0$, and $r = r_c$, which represents a D-brane position. In reference \cite{boer}, $r_c$ is the AdS boundary, and as $r_c \rightarrow \infty$, the string mass $m$ must be regularized. In this work we are going to take $r_c << l$, where $l$ is the AdS radius. This is a natural condition when it comes to fundamental strings.  In the background  defined by equations (\ref{BTZV}) and (\ref{h}), the string action is written as

\begin{equation}
    S= S_{NB}+ S_{B}
\end{equation}

\noindent where the first term is the Nambu Goto action and $S_{B}$ is the coupling of the string with the Kalb-Ramond field. For  $H_{\mu\nu\rho}$ defined in (\ref{h}), the only non zero component of the $B_{\mu\nu}$ field is $B_{xt}=\frac{r^2}{l}$. We are going to use the static  gauge, where the worldsheet coordinates $x^a=(\tau,\sigma)$  are identified with the spacetime coordinates $t$ and $r$, and only the transversal modes $X^I= X^I(t,r)$ are dynamical. In this gauge, $S_B$ can be written as

\begin{equation}
    S_B= -\frac{1}{2\pi\alpha'}\int dtdr \epsilon^{ab}B_{\mu\nu}\partial_aX^{\mu}\partial_bX^{\nu}=
    -\frac{1}{\pi\alpha'}\int dtdr\frac{r^2}{l}\partial_rX
\end{equation}

\noindent It will become clear  that this term will not contribute to the equations of motion in the approximation we are using. 

We focus now on the Nambu Goto action. For general planar $D$-dimensional AdS black holes, the metric can always be written as follows

\begin{equation}
ds^2= g_{\mu\nu}dx^{\mu}dx^{\nu}= g_{ab}dx^{a}dx^{b} + G_{IJ}dX^{I}dX^{J} \:, 
\label{genmetric}
\end{equation}

\noindent where both $g_{ab}$ and $G_{IJ}$ are independent of $X^I$ and $x^{a}= t,r$. In the static gauge, the Nambu-Goto action can be written as

\begin{equation}
S_{NG}= -\frac{1}{2\pi\alpha'}\int dt dr \sqrt{\det\gamma_{ab}} \:,
\end{equation}

\noindent where the induced metric is $\gamma_{ab}= G_{\mu\nu}\partial_{a}X^{\mu}\partial_{b}X^{\nu}$. The determinant can be written as

\begin{eqnarray}
\det(\gamma_{ab}) = \left| \begin{array}{cc}
G_{\mu \nu} \frac{d{X^{\mu}}}{d\tau} \frac{dX^{\nu}}{ d\tau} & G_{\mu \nu} \frac{dX^{\mu}}{d\tau}  \frac{d X^{\nu}}{d\sigma}\\
G_{\mu \nu} \frac{dX^{\mu}}{d\sigma}\frac{d X^{\nu}}{d \tau}& G_{\mu \nu}\frac{dX^{\mu}}{d\sigma} \frac{d X^{\nu}}{d\sigma} \end{array} \right|
= \left| \begin{array}{cc}
g_{tt}+G_{xx}\dot{x}^2  & G_{xx}\dot{x}x^{\prime}\\ 
G_{xx}x^{\prime}\dot{x}& g_{rr}+G_{xx}{x^{\prime}}^2 \end{array} \right|
\end{eqnarray}

\noindent where

\begin{eqnarray}
g_{tt}&=&-\frac{(r^2-m(t))}{l^2} \:, \nonumber \\
g_{rr}&=&\frac{l^2}{r^2-m(t)} \:,\nonumber \\
G_{XX}&=&\frac{r^2}{l^2} \:,
\label{gnovo}
\end{eqnarray}

By expanding the Nambu-Goto action, we get a power series of $\partial_{t}X^I$ and $\partial_{r}X^I$, which produces worldsheet interactions. As we are going to study only small fluctuations of the equilibrium value $X^I=0$, only quadratic terms in the action are considered. Since this approximation implies the regime of small velocities $|\partial_{t}X^I| <<1 $, we are in fact  taking the non-relativistic limit. In the quadratic approximation, the Nambu-Goto action can be written as

\begin{eqnarray}
S_{NG} &\approx& -\frac{1}{2\pi\alpha'}\int dr dt \sqrt{g(r)}g^{\mu\nu}(r)G_{IJ}(r)\partial_{\mu}X^{I}\partial_{\nu}X^{J} \:, \nonumber \\
&\approx&-\frac{1}{2\pi\alpha'}\int dt dr \left(\frac{r^4}{2l^4}F(r,t){X'}^2-\frac{\dot{X}^2}{F(r,t)} \right) \:,
\label{ngquadratic}
\end{eqnarray}

\noindent where 

\begin{eqnarray}
F(r,t) &=& 1-\frac{m(t)}{r^2} 
\end{eqnarray}

\noindent and $g(r)= \det g_{\mu\nu}$. In the second approximation equality we also dropped the constant term that does not depend on $X$. The equation of motion is

\begin{equation}
-\partial_t\left(\frac{\dot{X}}{F(r,t)}\right)+\partial_r\left(\frac{r^4}{l^4}F(r,t)X\right)+ \frac{r}{l}=0 \:.
\label{eqm1}
\end{equation}

\noindent where the last term comes from $S_B$. Let's define the dimensionless quantity 

\begin{equation}
\rho = \frac{r}{r_h}\:, \:\:\: G(\rho,t)= \rho^2-m(t) \:,\label{G}
\end{equation}

\noindent In terms of $\rho$, the horizon position  at $t= 0$ is located at $\rho=1$ and  the equation of motion becomes

\begin{equation}
-\frac{r_h^2}{l^4}\frac{\partial}{\partial \rho}\left[\rho^2G(\rho,t)X'\right] +\rho^2 \frac{\partial}{\partial t}\left[\frac{\dot{X}}{{G(\rho,t)}}\right] +\frac{r_h}{l}\rho=0 \:,
\end{equation}

\noindent where

\begin{equation}
\dot{X}=\frac{\partial X}{\partial t}\:, \:\:\: X'=\frac{\partial X}{\partial \rho} \:.\label{eqm}
\end{equation}

\noindent In the next section we are going to set the model for $m(t)$ and find an appropriate approximation to solve the equation of motion.

\section{The approximation and boundary condition}

A usual model concerning Vaidya black holes is $m(t)= 1+\tanh\left(at\right)$ so that, when $t\rightarrow 0$, the spacetime is a BTZ black hole with horizon $r_h$, and the open string is in thermodynamic equilibrium with Hawking radiation at temperature $T_0=\displaystyle\frac{r_h}{2\pi l^2}$. Then, the black hole expands and the system reaches another equilibrium point at $t\rightarrow \infty$ with $T_f=\displaystyle\frac{r_h}{\pi l^2}$. Note that, if we take $t\rightarrow  \infty$, we also get an AdS spacetime.  We are going to use a simplification and take a linear model such that $m(t) = 1+at$. In this model, the adiabatic approximation implies $a<<1$. Actually, we are studying the adiabatic expansion of the black hole close to $t=0$. 

Even for the simplest model $m(t) = 1+at$, the equation of motion (\ref{eqm}) does not have analytical solution. However, as the string is stretched between $\rho = \rho_c = 2$ and $\rho =1$ (the horizon position at $t=0$), we can try to use the approximation $\displaystyle\frac{r_h\rho}{l}<<1$ in order to solve the equation of motion. Nevertheless, in this time dependent background we shouldn't expect to find a solution like $X(\rho,t)= e^{i\omega t}X(\rho)$ for some frequency $\omega$.  Based on the solution of the equations of motion of the time independent BTZ model presented in \cite{boer}, the following solution is proposed

\begin{equation}
X= Y(\rho,t)\frac{\left[e^{i\rho\frac{l^2\omega}{r_h}}+ B e^{-i\rho\frac{l^2\omega}{r_h}}\right]}{\rho}
\end{equation}

\noindent where $\omega$ has frequency dimension and $B$ is a constant to be fixed by the boundary conditions. We realize that for $Y(\rho,t)= G(\rho,t)H(u)$, where $u=(G^2\omega/a)$, we get the following equation

\begin{equation}
u^2\partial^2_uH(u)+u\partial_uH(u)+\left(\frac{1}{4}u^2-\frac{1}{4}\right)H(u) + O\left(\frac{r_h}{l}\right)=0 \label{Bessel}
\end{equation}

\noindent where $O\left(\displaystyle\frac{r_h}{l}\right)$ represents higher orders in $\displaystyle\frac{r_h}{l}$. The equation (\ref{Bessel}) is just the Bessel equation of order $1/2$ for $H(u)$. At leading order in approximation $\displaystyle\frac{r_h}{l} << 1$, a solution of equation (\ref{eqm}) is

\begin{equation}
X(\rho , t) = A\: G(\rho,t) H_{1/2}\left(\frac{G^2 \omega}{a}\right)\frac{\left[e^{i\pi\nu \rho}+Be^{-i\pi\nu \rho}\right]}{\rho} +cc \label{sol}
\end{equation} 

\noindent where $H_{1/2}\left(\displaystyle\frac{G^2 \omega}{a}\right)$ is a Hankel function of order $1/2$ and $A$ is a normalization constant. The dimensionless frequency $\nu$ and the dimensional one $\omega$ are related by the expression

\begin{equation}
 \nu = \frac{l^2\omega}{\pi r_h}
 \label{freq}
\end{equation}

\subsection{Boundary Condition}

Now we are going to fix the constant $B$ by imposing boundary conditions.  The idea is to have a fundamental string attached to two D-branes, where the position of the first D-brane coincides with the position of the horizon at $t = 0$. In general the $B$ constant implies that a mode is reflected at $\rho = \rho_c$ and falls back into the horizon with phase shift. 
Let us start this section discussing the possibility of imposing Neumann boundary conditions to the equation of motion. By defining $z = \displaystyle\frac{G^2\omega}{a}$ and using the following relations of Bessel functions

\begin{eqnarray}
H_{n}(z) = J_{n}(z) + i Y_{n}(z) \:,  \nonumber\\
J_{1/2}(z) = \sqrt{\frac{2}{\pi z}} \sin z \:, \:\:\: J_{-1/2}(z) = \sqrt{\frac{2}{\pi z}} \cos z \:, \nonumber\\
Y_{n+1/2}(z) = (-1)^{n-1}J_{-n-1/2}(z) \label{H12}
\end{eqnarray}

\noindent we get 

\begin{eqnarray}
\frac{\partial X}{\partial \rho}\Big|_{\rho=\rho_c}= 2i\frac{\omega}{a} G(\rho_c,t)e^{iz(\rho_c,t)}\left( e^{i\pi\rho_c \nu}+ Be^{-i\pi\rho_c \nu}\right) \nonumber \\
+ e^{iz(\rho_c,t)}\frac{\partial}{\partial \rho}\left[\frac{e^{i\pi\rho_\nu}+ Be^{-i\pi\rho_ \nu}}{\rho}\right]_{\rho=\rho_c}+cc
\end{eqnarray}

\noindent Owing to time dependence of the first term, we can see that the solution (\ref{sol}) does not support a Neumann-type boundary condition.   
  So we choose $ \rho \in [1,2]$ and impose the following Dirichlet boundary conditions 

\begin{equation}
X(\rho,t)|_{\rho=1}=0,\:\:\: X(\rho,t)|_{\rho=2}=0
\end{equation}

\noindent We find that $B=-1$ and the frequencies are discretized.
Therefore, we have the following picture: the two string ends are fixed and, at $t=0$, one of then is fixed at the initial horizon position; as time runs, the horizon adiabatically crosses the string and we have a situation where the string connects the exterior and the interior of the black hole. Taking into account the properties of Bessel functions (\ref{H12}), it is easy to see that the worldsheet field is well-defined at the time dependent horizon ($\rho = at$).

\section{Quantization}

In this section, we proceed our analysis with the quantization of the string. The canonical commutation relations for the theory (\ref{ngquadratic}) are given by

\begin{eqnarray}
[X (r), X (r')]_{\Sigma} = 0,\:\:\:
 [X(r),n^{\mu} \partial_{\mu} X(r')]_{\Sigma} = \frac{2i\pi\alpha'}{\sqrt{h}} G_{XX} \delta (r-r'), \nonumber\\
\:[ n^{\mu} \partial_{\mu} X (r) ,n^{\nu} \partial_{\nu} X (r')]_{\Sigma} = 0.
\label{CCR}
\end{eqnarray}

\noindent where $\Sigma$ stands for a Cauchy surface in the $x^\mu= t,r$ part of the spacetime (\ref{genmetric}), $h_{ij}$ is the metric on $\Sigma$ induced from $g_{\mu\nu}$, and $n^{\mu}$ is the future-pointing unit normal to $\Sigma$. The worldsheet field is written as

 \begin{equation}
 X(\rho,t) = \sum_{\nu = 1}^{\infty} \frac{1}{\sqrt{\nu}} [a_{\nu} u_{\nu} (\rho, t) + a^{\dagger}_{\nu} u^{*}_{\nu} (\rho, t)]
  \label{sol1}
\end{equation}

\noindent with the normalized solutions 

\begin{equation}
u_{\nu}(\rho,t)= i \sqrt{\frac{\pi\alpha'}{r_h a}} G(\rho,t) H_{1/2}\left(\frac{G^2 \omega}{a}\right)\frac{
\sin(\pi\nu\rho)}{\rho} 
\label{u}
\end{equation}

\noindent and $u^*_{\nu}$ being the complex conjugate of $u_{\nu}$. For functions $f^I(x),g^I(x)$ satisfying the equation of motion, we can define the following inner product\footnote{It can be shown that this inner product is independent of the choice of $\Sigma$, just as the standard Klein--Gordon inner product \cite{Birrell:1982ix}.}

\begin{equation}
 (f,g)_\Sigma=-\left(\frac{i}{2\pi\alpha}\int_\Sigma d\rho \sqrt{h}\, n^\mu
 G_{IJ}(f^I  \partial_\mu g^{J*} - \partial_\mu f^I \,g^{J*}\right).
\label{innerprod}
\end{equation}

\noindent and it is not difficult to show that the canonical commutation
relations (\ref{CCR}) are equivalent to

\begin{equation}
 [(f,X)_\Sigma,(g,X)_\Sigma]_\Sigma=(f,g^*)_\Sigma
 \label{CCRcond_innerprod}
\end{equation}
 
 \noindent $\forall f,g$ satisfying the equation of motion (\ref{eqm}).

As the non zero components of $g_{\mu\nu}$ and $G_{IJ}$ are given by (\ref{gnovo}), we can choose 

\begin{equation}
n^\mu=((-g_{tt})^{-1/2},0,0),
\label{vetor}
\end{equation}

\noindent such that $g_{\mu\nu}u^\mu u^\nu =-1$. The induced metric on $\Sigma$ from $g_{\mu\nu}$ is just 

\begin{equation}
h_{\rho \rho}=g_{\rho\rho} 
\label{mi}
\end{equation}

\noindent given in equation (\ref{gnovo}). Hence, for this model, the inner product for $u_{\nu}(\rho,t)$ and $u_{\nu}^*(\rho,t)$ is, up to normalization constants,

\begin{equation}
(u_{\nu},u_{\nu'}^*)_{\Sigma} = i \int_{\Sigma}d\rho \cos\pi\rho(\nu-\nu') W(t) 
\label{uprod}
\end{equation} 

\noindent where $W(t)$ is the wronskian

\begin{equation}
W(t)= \frac{ \omega G^2 (\rho,t)}{a} \left[H_{1/2} (\rho,t) \partial_t H_{1/2}^* (\rho,t)- \partial_t H_{1/2} (\rho,t) H_{1/2}^* (\rho,t)\right] \:,
\end{equation}

\noindent and we are using the notation $H_{1/2}\left(\displaystyle\frac{G^2 \omega}{a}\right)\rightarrow H_{1/2}\left(\rho,t\right)$. Note that the term $G^2(\rho,t)$ comes from the induced metric and the unitary vector $n^{\mu}$. The inner product in equation (\ref{uprod}) seems time dependent, but using properties of the Hankel functions one can see that
\begin{equation}
[H_{1/2} (\rho,t) \partial_t H_{1/2}^* (\rho,t)- \partial_t H_{1/2} (\rho,t) H_{1/2}^* (\rho,t)= \frac{-i a}{\omega G^2(\rho,t)}
\end{equation}

\noindent so

\begin{equation}
 (u_\nu ,u^*_{\nu'})_\Sigma= \delta(\nu-\nu') \:.
 \label{33}
\end{equation}

\noindent Therefore, it is straightforward to realize that the condition (\ref{33}) implies the canonical commutation relations (\ref{CCR}) and

\begin{eqnarray}
 [a_\nu ,a_{\nu'}^{\dagger}]= \nu \delta(\nu - \nu'),\:\:\: [a_\nu,a_{\nu'}]=0 \:, \:\:\: [a^{\dagger}_\nu,a^{\dagger}_{\nu'}]=0 \:.
 \label{oscila}
\end{eqnarray}

Once we have chosen the Cauchy surface $\Sigma$ and the vector $n^{\mu}$, we can define the vacuum associated to $\Sigma$: 

\begin{equation}
a_\nu|0\rangle_{\Sigma}=0 \:.
\label{vac}
\end{equation}

\noindent Owing to the time dependence of the induced worldsheet, it is expected that there will be no string mode conservation and the vacuum is not unique.  Although we are not studying here string mode creation, we are going to show in the next section that the equal times two-point function will depend on time, which means that will depend on $\Sigma$ and consequently on the choice of the vacuum.

\section{Two-point function}

For metrics with timelike Killing vectors, the equal times worldsheet two-point function does not depend on time. Let us show that this is not the case for the model studied in this article. Considering equations (\ref{sol1}), (\ref{u}) and (\ref{oscila}), we are going to compute the two-point function at equal times firstly at zero temperature limit, which is defined by 

\begin{eqnarray}
<0|X(\rho, t) X(\rho', t)|0> = \nonumber\\
= \sum^{\infty}_{\nu =1} \frac{1}{\sqrt{\nu}}\sum_{\xi =1}^{\infty} \frac{1}{\sqrt{\xi}} <0|[a_{\nu} u_{\nu}(\rho,t) + a^{\dagger}_{\nu} u^{*}_{\nu} (\rho,t)][a_{\xi} u_{\xi}(\rho,t) + a^{\dagger}_{\xi} u^{*}_{\xi} (\rho,t)]|0> \:.
\end{eqnarray}

\noindent Here the vacuum is the one defined in (\ref{vac}) and the ``equal times" means that $\rho$ and $\rho'$ belongs to the same Cauchy surface $\Sigma$. After some manipulation of the expression given above, we have

\begin{eqnarray}
<0|X(\rho, t) X(\rho', t)|0> 
= \frac{\pi\alpha'}{4 r_h a} \sum^{\infty}_{\nu =1} \frac{G(\rho, t) G(\rho',t)}{\rho\rho'} H_{1/2} (z) H^*_{1/2} (z')\times \nonumber\\
 \times [e^{i\nu (\rho - \rho')} - e^{i\nu (\rho + \rho')} - e^{-i\nu (\rho + \rho')} + e^{-i\nu (\rho - \rho')}]
 \label{2pontos}
\end{eqnarray}

\noindent where $z = G^2 \omega/a$. Using properties (\ref{H12}), we get

\begin{eqnarray}
<0|X(\rho, t) X(\rho', t)|0> 
= \frac{\alpha' l^2}{2 r^2_h} \frac{1}{\rho\rho'} \sum_{\nu =1}^{\infty} \frac{(e^{i\nu M}  -e^{i\nu N} -e^{i\nu P} + e^{i\nu Q})}{\nu}
\label{2pontos2}
\end{eqnarray}

\noindent with the following definitions:

\begin{eqnarray}
M = (\rho-\rho')\left[\left(\frac{r_h}{al^2}\right)(\rho +\rho')(\rho^2 + \rho^{'2} - 2m) +1\right] \nonumber\\
N = (\rho+\rho')\left[\left(\frac{r_h}{al^2}\right)(\rho -\rho')(\rho^2 + \rho^{'2} - 2m) +1\right] \nonumber\\
P = (\rho+\rho')\left[\left(\frac{r_h}{al^2}\right)(\rho -\rho')(\rho^2 + \rho^{'2} - 2m) -1\right]\nonumber\\
Q = (\rho-\rho')\left[\left(\frac{r_h}{al^2}\right)(\rho +\rho')(\rho^2 + \rho^{'2} - 2m) -1\right]
\end{eqnarray}

\noindent Finally, after solving the sum, the expression (\ref{2pontos2}) becomes

\begin{eqnarray}
<0|X(\rho, t) X(\rho', t)|0> 
= -\frac{\alpha' l^2}{2 r^2_h} \frac{1}{\rho\rho'} \ln [(1 - e^{iM})(1 - e^{iN})(1 - e^{iP})(1 - e^{iQ})]
\end{eqnarray}

\noindent Note that, as $\rho \approx \rho'$, the equal times two-point function has the same kind of singularity as the flat space one. However, it is time dependent, which means that the behaviour of the two-point function depends on worldsheet $\Sigma$ surface which $\rho$ and $\rho'$ belongs to. This is a consequence of vacuum dependence on $\Sigma$.
 
Let us now move on to incorporate finite temperature effects. As it was shown in \cite{Lawrence:1993sg, Frolov:2000kx}, near the horizon at $t=0$ the worldsheet action is the same as that of a Klein–Gordon field near a two-dimensional black hole. Then, the string modes are thermally excited, presenting a black-body spectrum determined by the Hawking temperature. However, in our time dependent model, we cannot naively construct a thermal density matrix defined by (where $H$ is the time dependent Hamiltonian of the system)

\begin{equation}
\rho_H=\frac{1}{Z}e^{-\beta H} \:,
\label{rhoh}
\end{equation}

\noindent just because this density matrix does not satisfy the quantum Liouville-von Neumann (LvN) equation, and it is difficult to relate $1/\beta$ to the equilibrium temperature. Following the references \cite{Kim:2000xb,Kim:2001pg,Dodonov}, one can study the evolution of non equilibrium systems with respect to invariants of the system. By invariant one means an operator $I$ defined on a Cauchy surface $\Sigma$
which satisfies the LvN equation, that is, $\displaystyle\frac{d I}{d t}=0$. Given the invariant $I$, one can introduce an analogue of the thermal density operator 

\begin{equation}
\rho_I=\frac{e^{-\beta I}}{Tr e^{-\beta I}}\;\;.
\end{equation}

\noindent that satisfies the LvN equation\footnote{The LvN approach was used in the context of string theory in \cite{Marchioro:2020qub} and \cite{Blau:2004cm}.} and $Tr\hat\rho_I=1$. A convenient choice of invariant for the system described here is $I=H(t=0)$ because it reduces to the standard choice in the case of a time independent system, and just at $t=0$ the system is at thermal equilibrium with the BTZ Hawking temperature $T_0=\beta^{-1}= \displaystyle\frac{r_h}{2\pi l^2} $.  Note that this is also consistent with the adiabatic approximation provided
that $H(t)$ varies sufficiently slowly with time near $t=0$.

 At $t=0$, the system is an open string attached between the BTZ horizon and a brane. So, the invariant thermal density is the one described in \cite{boer}, given by
 
\begin{equation}
\rho_I=\frac{e^{-\beta H(0)}}{Tr e^{-\beta H(0)}},\;\; H(0)=\sum_{\omega>0}\omega a^{\dagger}_{\omega}a_{\omega}  .
 \end{equation}
 
\noindent With this density matrix we can define thermal two-point function

\begin{equation}
\langle X(\rho,t)X(\rho',t)\rangle_{\beta}= Tr(\rho_I X(\rho,t)X(\rho',t))
\end{equation}
 
\noindent which can be written as

 \begin{eqnarray}
 <|X(\rho, t) X(\rho', t)>_{\beta} &=&<0|X(\rho, t) X(\rho', t)|0> \nonumber \\
  &+& 2\frac{\alpha' l^2}{2 r^2_h} \frac{1}{\rho\rho'} \sum_{\nu =1}^{\infty} \frac{1}{e^{\beta \omega }-1}\frac{(e^{i\nu M} - e^{i\nu N} -e^{i\nu P} + e^{i\nu Q})}{\nu}
 \end{eqnarray}
 
 As usual in real time finite temperature theories, the two-point function is written as a sum of the zero temperature function and the finite temperature correction. Expressing the bosonic distribution as a geometric series and using $\beta \omega = \displaystyle\frac{\nu}{2\pi}$, we get
 
 \begin{eqnarray}
 <|X(\rho, t) X(\rho', t)>_{\beta} = <0|X(\rho, t) X(\rho', t)|0>\nonumber \\
 -\frac{\alpha' l^2}{2 r^2_h} \frac{1}{\rho\rho'}  \sum_{l=0}\ln [(1 - e^{iM-\frac{1+l}{2\pi}})(1 - e^{iN-\frac{1+l}{2\pi}})(1 - e^{iP-\frac{1+l}{2\pi}})(1 - e^{iQ-\frac{1+l}{2\pi}})]
 \end{eqnarray}
 
\noindent The new series can be resolved in terms of theta ($\vartheta (q)$) and q-Gamma functions ($\Gamma_q(x)$). Defining
 
\begin{eqnarray}
m(q)&=&\left(\frac{\vartheta_2(q)}{\vartheta_3(q)}\right)^4\nonumber \\  
K(q)&=&\frac{\pi}{2}\vartheta_3(q)^2 \nonumber \\
 q&=& e^{-\beta},\:\:x= 1-i\frac{y}{2\pi},
 \end{eqnarray}

\noindent and the function 

\begin{eqnarray}
F(x)=\sum_{l=0}\ln(1+q^{x+l})&=&(1+x)\ln(1-q)+\frac{1}{24}\ln\left(\frac{m(q)(1-m(q))^4}{16q}\right)+\frac{1}{2}\ln\left(\frac{2K(q)}{\pi}\right) \nonumber \\
&-&\ln\Gamma_q(x) \label{sum}
\end{eqnarray}

\noindent the finite temperature two-point function can be rewritten as
  
 \begin{eqnarray}
 <|X(\rho, t) X(\rho', t)>_{\beta} &=&<0|X(\rho, t) X(\rho', t)|0>\nonumber \\
 &-&\frac{\alpha' l^2}{2 r^2_h} \frac{1}{\rho\rho'}\left[F( M)+F(N)+F(P)+F(Q)\right]
 \end{eqnarray}

\noindent where q-Gamma $\Gamma_q(x)$ function is defined from the q-factorial $(a;q)_\infty=\displaystyle\prod_{k=1}^{\infty}(1-aq^k),|q|<1$,

\begin{equation}
\Gamma_q(x)= \frac{(q;q)_\infty(1-q)^{1-x}}{(q^x;x)_\infty} \: ,
\end{equation}

\noindent and the theta functions are defined by the usual product

\begin{equation}
\vartheta_2(q)=2q^{1/8}\prod_{n=1}^{\infty}(1-q^n)(1+q)^2,\:\:  \vartheta_3(q)=2q^{1/8}\prod_{n=1}^{\infty}(1-q^n)(1+q^{n+1/2})^2
\end{equation}

 Using the relation

\begin{equation}
\Gamma_q(x) =\frac{1-q}{1-q^x}\Gamma_q(x+1) \,,
\end{equation}

\noindent and $\vartheta$-function properties, we can see that the short-distance behaviour of finite temperature two-point function is the same as the zero temperature one, as usual in quantum field theory. In particular, the only singular terms come from the zero temperature part.
 

\section{Conclusion}

In this letter, the canonical quantization of the open string defined in a time dependent BTZ black hole is carried on. The boundary conditions are such that the string is attached between two branes, with the position of one of them coinciding with the position of the horizon at $t=0$. So, after $t=0$, the endpoints of the open string connect points inside and outside the horizon. This kind of configuration was studied in reference \cite{BatoniAbdalla:2007zv} for the Rindler space, where its relation with the “stretched horizon” model was explored \cite{suskind1,suskind2}.

Due to the temporal dependence of the induced metric, the vacuum depends on the Cauchy surface (where the worldsheet fields are defined), orthogonal to a future-pointing unit vector. This becomes clear when calculating equal times two-point function. Although the short distance behaviour is the same as that observed in flat space, the equal times two-point function is time dependent, which implies that it depends on the Cauchy surface. As mentioned earlier, this is a consequence of the vacuum not being globally defined in the worldsheet. As an application of this work, it would be important to investigate the scenario studied in \cite{boer}  and verify the effects of temporal dependence of the geometry on Brownian motion at the boundary of the AdS space. To this end, it will be necessary to solve the equation of motion beyond the approximation $\displaystyle\frac{r_h}{l}<<1$.

\end{document}